\documentclass[twocolumn,aps,prb,showpacs]{revtex4-1}
\usepackage[latin9]{inputenc}
\usepackage{xcolor}
\usepackage{units}
\usepackage{textcomp}
\usepackage{amsmath}
\usepackage{amssymb}
\usepackage{graphicx}

\makeatletter

\usepackage[singlelinecheck = false, justification=raggedright]{caption}
\@ifundefined{showcaptionsetup}{}{
\PassOptionsToPackage{caption=false}{subfig}}
\usepackage{subfig}
\makeatother

\begin{document}

\title{Carrier Density and Thickness Dependent Proximity Effect in Doped Topological Insulator - Metallic
Ferromagnet Bilayers}

\author{Yaron Jarach, Gad Koren, Netanel H. Lindner , Amit Kanigel}

\affiliation{Physics Department, Technion-Israel Institute of Technology, Haifa
3200003, Israel.}

\begin{abstract}
We use magneto-conductivity to study the magnetic proximity effect on surface states of doped topological insulators. Our bilayers consist of a layer of Fe$_7$Se$_8$, which is a metallic ferrimagnet and a layer of Bi$_{0.8}$Sb$_{1.2}$Te$_{3}$ which is a highly hole-doped topological insulator. Using transport measurements and a modified Hikami-Larkin-Nagaoka model, we show that the ferromagnet shortens significantly the effective coherence length of the surface states, suggesting that a gap is opened at the Dirac point. 
We show that the magnetically induced gap persists on surface states which are separated from the magnet by a topological insulator layer as thick as 170 [nm]. Furthermore, the size of the gap is found to be proportional to the magnetization that we extract from the anomalous Hall effect.
Our results give information on the ties between carrier density, induced magnetization and magnetically induced gap in topological insulator/ferromagnet bilayers. This information is important both for a  theoretical understanding of magnetic interactions in topological insulators and for the practical fabrication of such bilayers, which are the basis of various suggested technologies, such as spintronic devices, far infra-red detectors etc.

\end{abstract}

\maketitle

\section{Introduction}

Strong topological insulators (TIs) are band insulators with time reversal
symmetry protected gapless surface states\cite{Bernevig2013,Hasan2010,Zhang2009}.
The dispersion of the surface states exhibits an odd number of massless Dirac fermions and a strong spin-momentum locking\cite{Bernevig2013,Zhang2009,Miyamoto2012,Kuroda2016}. Magnetic fields normal to the surface will
open a gap at the Dirac
point \cite{Bernevig2013,Hasan2010}. Magnetically gapped surface states host a variety of fascinating phenomena,
including the quantum anomalous Hall effect (QAHE)\cite{Okuyama2019,Lee2018}.

One way to break time reversal
symmetry is by interfacing a TI with a ferromagnetic
(FM) material \cite{Okuyama2019,Lee2018,Otrokov2017,Lu2014,Lu2011,He2017,Katmis2016,Wei2013,Lang2014}.
The surface states interact with the magnetic moments and acquire a magnetic proximity effect induced energy gap\cite{Okuyama2019,Lee2018,Wei2013}. One major advantage of this method is the ability
to pattern the FM layer and create spatially varying magnetic interactions\cite{Lindner}.

Magnetic proximity effect induced energy gaps on surface states in TI/FM bilayers may result for two different mechanisms\cite{Li2015}: 1) induction of spin polarization
in the TI. 2) exchange mechanisms such as RKKY that 
couple the surface states and the magnetic material leaving the TI non-magnetic
\cite{Li2015}. Polarized neutron reflection and magnetic
second harmonic generation measurements show that
only 1-2 nanometers in the TI are magnetized by the magnetic interface\cite{Katmis2016,Li2017,Lee2016}. 
However, magnetic exchange coupling in FM/Metal structures can penetrate tens of nanometers
into the metal\cite{Li2015,Bruno1992}.

While magnetic proximity effect is known to gap the surface states on the interface that is in direct contact with the FM\cite{Okuyama2019}, an important question 
is whether magnetic exchange coupling can also gap surface states that are not in direct 
contact with the FM. This remains so far an open question\cite{Okuyama2019,Lee2018,Yang2013,Yang2019,Zheng2016,Lee2014}. 
In this work, we studied this question for FM/doped-TI bilayers, for which one of the surface states is in direct contact with the FM, while the opposing (exposed) surface is separated from the FM by the TI bulk.

Direct ways to probe the surface states dispersion are angle-resolved photoemission spectroscopy (ARPES) and scanning tunneling microscopy (STM). However,
these are surface methods that require high quality surfaces which
are difficult to achieve in TI/FM bilayers. Simpler methods to probe
the surface states energy gap involve electrical transport measurements. 
The anomalous Hall effect in the TI can indicate time reversal
symmetry breaking and
gapped surface states\cite{Yang2013,Tang2017}. Furthermore, the surface states gap 
strongly affects the magneto-conductivity\cite{Lu2014,Lu2011}.
The gap size can be calculated using a modified Hikami-Larkin-Nagaoka
(m-HLN) equation\cite{Lu2014,Lu2011}. We discuss this  equation in
section \ref{sec:Results-and-Discussion}.

Naturally, insulating TIs and FM are preferable to explore the surface states gap using transport measurements.
Thus, previous transport studies on TI/FM bilayers used low carrier
density $\left(n\lesssim10^{19} \:\left[cm^{-3}\right]\right)$ ultra-thin
$\left(t\lesssim10 \:\left[nm\right]\right)$ TIs and insulating FM \cite{Lee2018,Lu2014,Lu2011,He2017,Katmis2016,Wei2013,Lang2014,Li2017,Yang2013,Yang2019,Zheng2016,Tang2017,Kandala2013}.
However, low carrier density samples might lack the exchange mechanisms needed to open a gap in the exposed surface states, since exchange mechanisms are highly sensitive to carrier density\cite{Li2015,Bruno1992,Kim2011,He2011,Wang2016,Bansal2012}. We therefore used doped TIs in this study.

In addition, proximity to a FM metal can be considerably different from a proximity to a FM insulator. This is because the metal can modify the TI's band structure, chemical potential and the magnetic coupling mechanism\cite{Marmolejo-Tejada2017,Hsu2017,Li2015,Li2015a}. Therefore, in this study we focused on TI/FM metal bilayers. This is also of practical interest since there are only a few TI-compatible ferromagnetic insulators
and some possible TI/FM applications require a FM metal\cite{Mellnik2014}.

Here, we study the electrical transport properties of highly doped TI/FM-metal bilayers 
with varying TI thicknesses and carrier densities. By fitting the bilayers magneto-conductivity with the modified Hikami-Larkin-Nagaoka
model \cite{Lu2014,Lu2011} for surface states magneto-conductivity we extract the magnetic proximity effect induced energy gap. Our results indicate 
a long-range, carrier density dependent proximity effect which gaps both surface states, the 
one in direct contact with the FM but also the surface states on the exposed surface of the TI. We also verify the validity of m-HLN equation
in our system despite the large amount of bulk conducting channels.

\section{Results and Discussion\label{sec:Results-and-Discussion}}

\begin{figure}
\subfloat[]{\includegraphics[width=8.6cm]{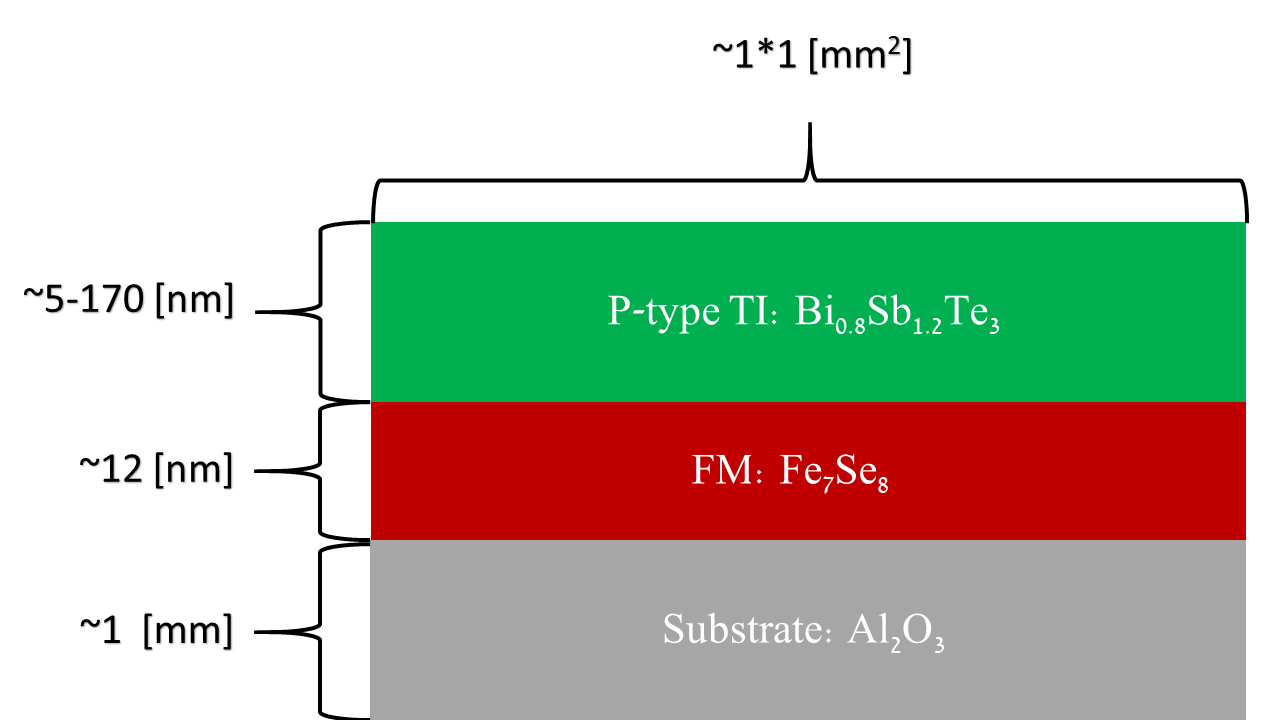}

}

\subfloat[]{\includegraphics[width=8.6cm]{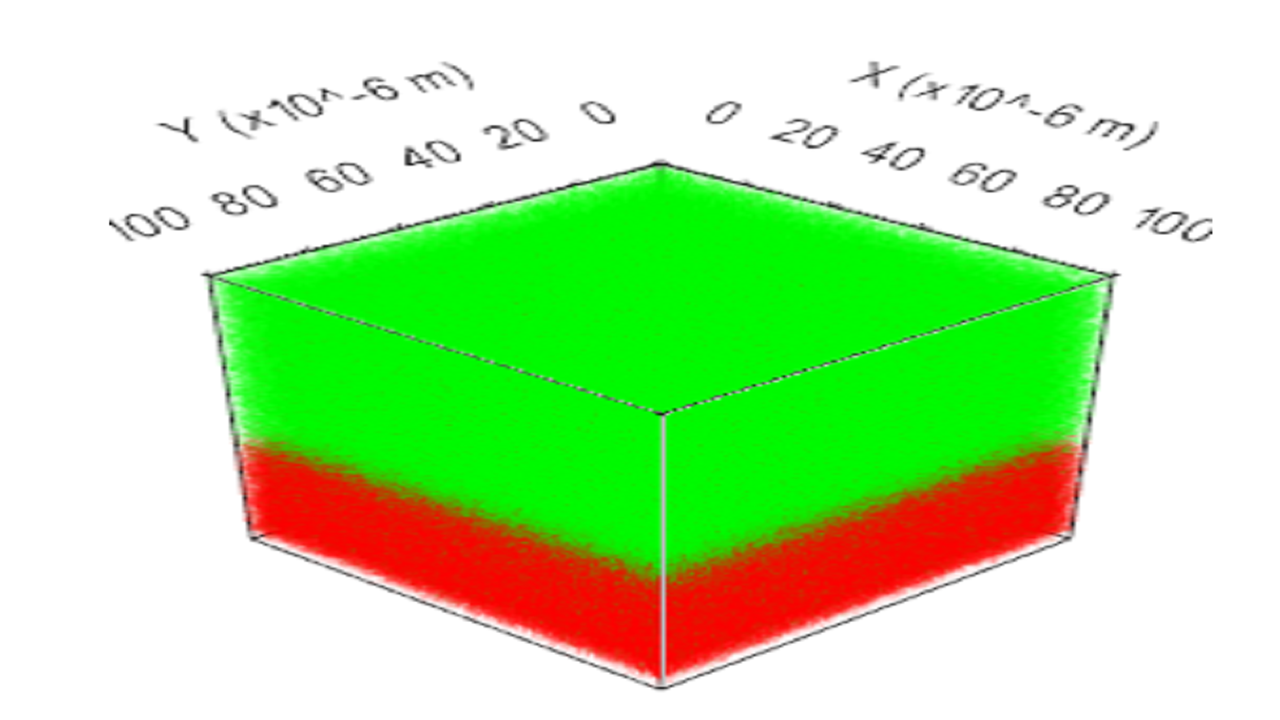}

}

\caption{(a) The bilayer scheme (b) A Time Of Flight Secondary Ion Mass Spectroscopy (TOF-SIMS) 3D image of Bi (green) Fe (red) atoms relative intensities in 170 [nm] $Bi_{0.8}Sb_{1.2}Te_{3}$ on 12 [nm]
$Fe_{7}Se_{8}$ on $Al_{2}O_{3}$. X and Y axes are spatial coordinates
and Z is etching time. The FM etching rate is $~5$ times
lower than the TI rate so the layer thicknesses are not scaled. The
bottom transparent layer is the $Al_{2}O_{3}$ substrate. The image
shows that the bilayer consists of two distinct phases with an interface
volume much smaller than the sample volume. \label{fig:Sample-scheme}}
\end{figure}

Figure \ref{fig:Sample-scheme} shows the structure of our bilayers.
The TI layer is a highly p-type doped $Bi_{0.8}Sb_{1.2}Te_{3}$ film
with thickness ranging from 17 [nm] to 170 [nm]. This compound has
an hexagonal lattice structure and a ~0.3 [eV] band gap\cite{Zhang2011,Weyrich2016}.
We note that for TIs thinner than 5 [nm] the surface states may be gapped due to
a hybridization between them\cite{Okuyama2019,Zhang2010}. As we show, the thickness variation
results also in carrier density variation, enabling us to study both
thickness and carrier density effects.

The FM layer is a $Fe_{7}Se_{8}$ film, which is a hexagonal metallic ferrimagnet with $T_{c}\sim450 \:\left[K\right]$ and a canted
easy magnetization axis\cite{Takemura1997,Terzieff1978}. We grew the bilayers
on $Al_{2}O_{3}$ substrates using pulsed laser deposition (PLD)
at 280 [C] and a pressure of $\sim1*10^{-7}\left [Torr\right]$.
X-ray diffraction shows that the FM and the TI grow along the c-axis, with grain size of about 45 [nm] (see appendix \ref{xray data appendix}).
The FM layer thickness was 12 [nm] in all of the bilayers. All transport measurements were done at T=1.5 [K] using the
standard 4-contact Van-der-Pauw (VDP) technique.

\textit{Surface states properties:} The $Fe_{7}Se_{8}$ film out-of-plane magnetization component opens
an energy gap ($\Delta$) at the Dirac node of the surface states dispersion.
In this case, the surface states magneto-conductivity (MC) obeys a modified Hikami-Larkin-Nagaoka
model\cite{Lu2014,Lu2011} (m-HLN).
We begin our analysis using the small-gap limit of the model (for details see appendix \ref{Modified-HLN-formula}):
\begin{align}
\Delta\sigma{\left(B\right)}= & -\dfrac{e^{2}}{\pi h}\alpha\left[\psi\left(\frac{1}{2}+\dfrac{l_{B}^{2}}{l_{\phi1}^{2}}\right)-ln\left(\dfrac{l_{B}^{2}}{l_{\phi1}^{2}}\right)\right]\label{eq:mHLN function}\\
l_{\phi1}^{-2} & \equiv l_{c}^{-2}+l_{e}^{-2}f_{\left(\widetilde{\Delta}\right)}\label{eq:lphi def}
\end{align}
Where $\Delta\sigma\left(B\right)\equiv\sigma\left(B\right)-\sigma\left(0\right)$ is the MC
of one surface and $\psi$ is the digamma function\cite{Lu2014,Lu2011}. The parameter
$l_{B}=\sqrt{\dfrac{\hbar}{4eB}}$ is the magnetic length. The length scale $l_{\phi1}$
is the effective coherence length of the surface states in the presence of an induced
gap. As indicated by Eq. \ref{eq:lphi def}, $l_{\phi1}$ depends on $l_{c}$, which is the surface states coherence length; 
$l_{e}$, which is the elastic mean free path; and $\widetilde{\Delta}\equiv{\Delta}/{\epsilon_{F}}$ is the normalized
magnetically induced energy gap. Furthermore, in Eq. \ref{eq:mHLN function} $\alpha$ and $f$ are functions of $\widetilde{\Delta}$, 
whose form is given in appendix \ref{Modified-HLN-formula} \cite{Lu2014,Lu2011}.

\begin{figure}
\begin{centering}
\includegraphics[width=8.6cm]{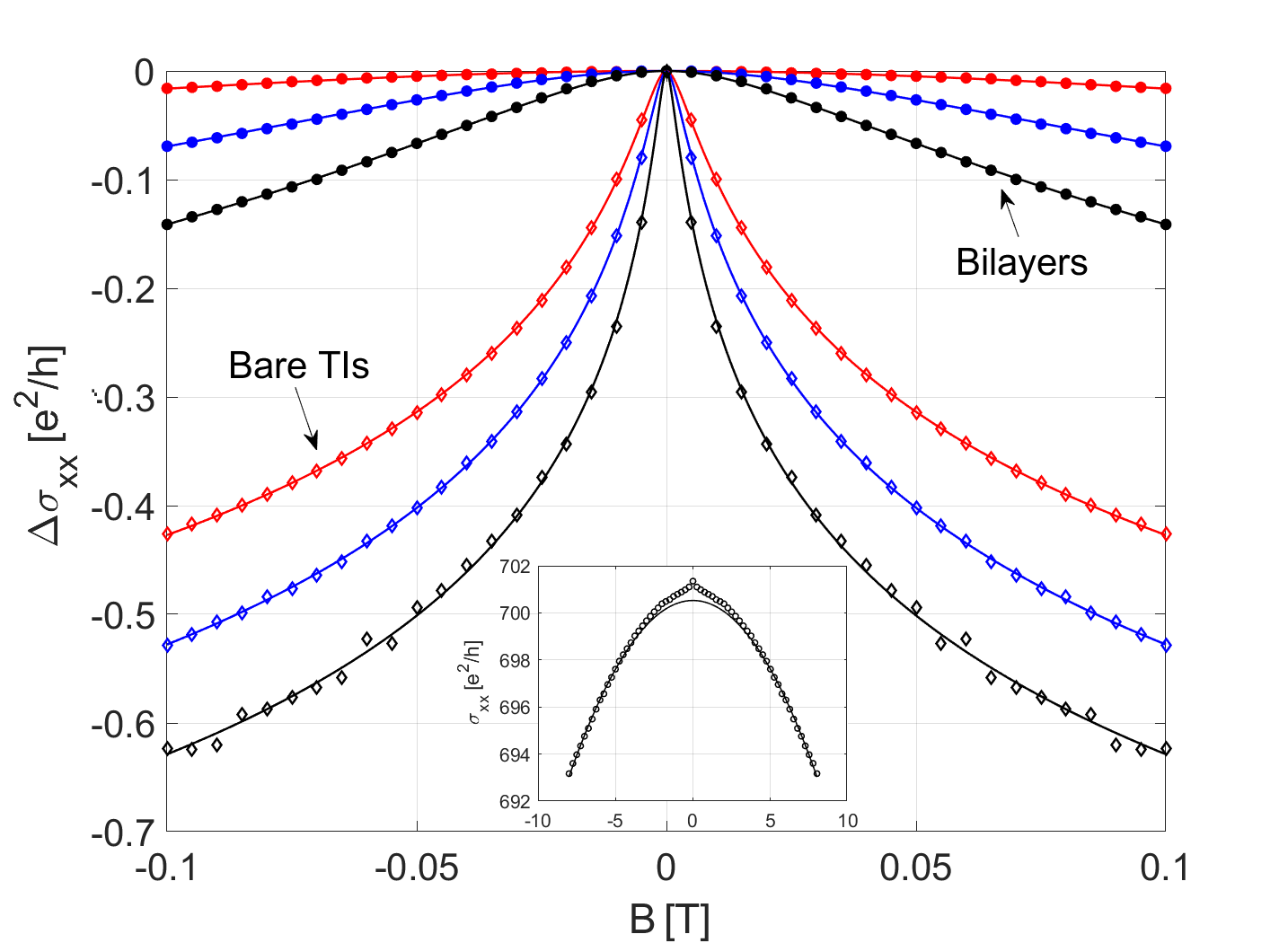} 
\par\end{centering}
\caption{Magneto conductivity (MC), $\Delta\sigma_{xx}=\sigma_{xx}\left(B\right)-\sigma_{xx}\left(0\right)$,
of bare TI films (diamonds) and TI/FM-metal bilayers (filled circles) with a m-HLN
model fit. The TI thicknesses are 17 [nm] (red), 73 [nm] (blue) and 170 [nm]
(black). (Inset) Conductivity vs magnetic field of a bilayer with 170 [nm]
TI, with a Kohler quadratic fit in the high fields. The fit gives
an upper limit on the TI bulk MC term, showing it is negligible at
low fields.\label{fig:MC data with fit and a bare-TI bilayer comp}}
\end{figure}

We can therefore study magnetic interactions in our bilayers by fitting
the MC signal with Eq. \ref{eq:mHLN function}. In Figure \ref{fig:MC data with fit and a bare-TI bilayer comp}
we show a comparison between the MC data of a bilayer and a bare
TI film for several different TI thicknesses. We fitted the data with the m-HLN model for the MC given by Eq\ref{eq:mHLN function}. 
Below we show that the bilayer MC data is dominated by the exposed surface states 
and not by other transport channels, and discuss the fitting parameters of Eq. \ref{eq:mHLN function}.
Afterwards we return to our main topic which is the extracted energy gap ($\widetilde{\Delta}$) properties.

\textit{Transport channels analysis:} Eq. \ref{eq:mHLN function} gives the conductivity of a system with a single 2D massive Dirac Fermion dispersion. 
The good quality of the fitting of Eq. \ref{eq:mHLN function} to our data demonstrates 
the surface states robustness even in proximity to a FM-metal 
interface \cite{Marmolejo-Tejada2017,Hsu2017}. It also indicates
that the MC of both the $Fe_{7}Se_{8}$ and the TI bulk are negligible. 

To verify that the $Fe_{7}Se_{8}$ MC is negligible
we had measured a bare $Fe_{7}Se_{8}$ film and found MC values
below $1\%$ of the total MC of the BLs. For the TI bulk, the MC can have different contributions. First, the classical Kohler equation\cite{Kim2011,He2011} term: $\sigma_{xx}^{bulk}(B)\sim\sigma_{xx}^{bulk}(0)\left(1-2\mu_{xx}^{2}B^{2}\right)$,
with $\mu_{xx}$ the mobility. By fitting this equation to the high-field
MC data and extrapolating down to zero field, we found that the 
this term is an order of magnitude smaller than the total MC of the bilayers.
The inset in Fig. \ref{fig:MC data with fit and a bare-TI bilayer comp}
shows this fit for a bilayer with 170 [nm] thick TI (thickest in this study).
From this fit we obtained a mobility of about 90 [cm$^{2}$/(v$\cdot$s)]. This mobility value is within $10\%$ from 
the value obtained by Hall measurements, which verifies our MC extrapolating procedure.

Previous studies have shown that in thin TI films the bulk MC could also have a weak anti-localization term due to strong spin-orbit coupling\cite{Kim2011, Steinberg2011}. This gives a term similar to Eq. \ref{eq:mHLN function} with $\alpha=1/2$\cite{Kim2011, Steinberg2011} and a zero magnetically induced energy gap. However, other studies have shown experimentally that the bulk weak anti-localization term is smaller than $20\%$ of the surface term\cite{He2011}, or even negligible\cite{Park2018,Li2019d}. Moreover, for 73-170 [nm] thick TIs we found lower bulk carrier density and higher bulk mobility in the bilayers than in the bare TI films (see below how we extracted these properties in the bilayers). The opposite behaviour is found for the thinner samples. The bulk coherence length is proportional to the mobility and is inversely proportional to the carrier density\cite{Kim2011}. The weak anti-localization term is stronger for higher coherence lengths (equation \ref{eq:mHLN function}). Therefore if the bulk term would be  significant in our samples, we would get a higher MC in the bilayers with 73-170 [nm] thick TI than in the bare TIs. This is clearly not the case in figure \ref{fig:MC data with fit and a bare-TI bilayer comp}, where the MC is much higher in the bare TIs for every TI thickness. We therefore conclude that we can neglect the bulk anti-localization term in our samples. 

Although a TI layer has two surfaces and sometimes also a bulk channel as mentioned above, the MC data can be fitted using Eq. \ref{eq:mHLN function} which corresponds to a single surface. This was reported in several previous works \cite{Kim2011,Wang2016,Bansal2012,He2011,Zhang2012,Steinberg2011,Park2018,Li2019d}. The amplitude however may vary from the theoretically predicted of $\alpha=1/2$, depending on sample properties\cite{Kim2011,Wang2016,Bansal2012,He2011,Zhang2012,Steinberg2011,Park2018,Li2019d}. One common explanation is that the two surfaces (and the bulk) can couple even for TI thickness of $\sim$100 [nm]\cite{Kim2011,Wang2016,Bansal2012,Steinberg2011,Park2018,Li2019d}. This results in $\alpha=1/2$ when all channels are coupled, or $\alpha=1/2$ per channel when the channels are fully decoupled\cite{Steinberg2011,Park2018}. In our samples we got $\alpha\sim~0.6-0.7$ for all thicknesses, both in the bare TIs and bilayers. 

\begin{figure}
\subfloat[\label{fig:fitted lphi1}]{\includegraphics[width=8.6cm]{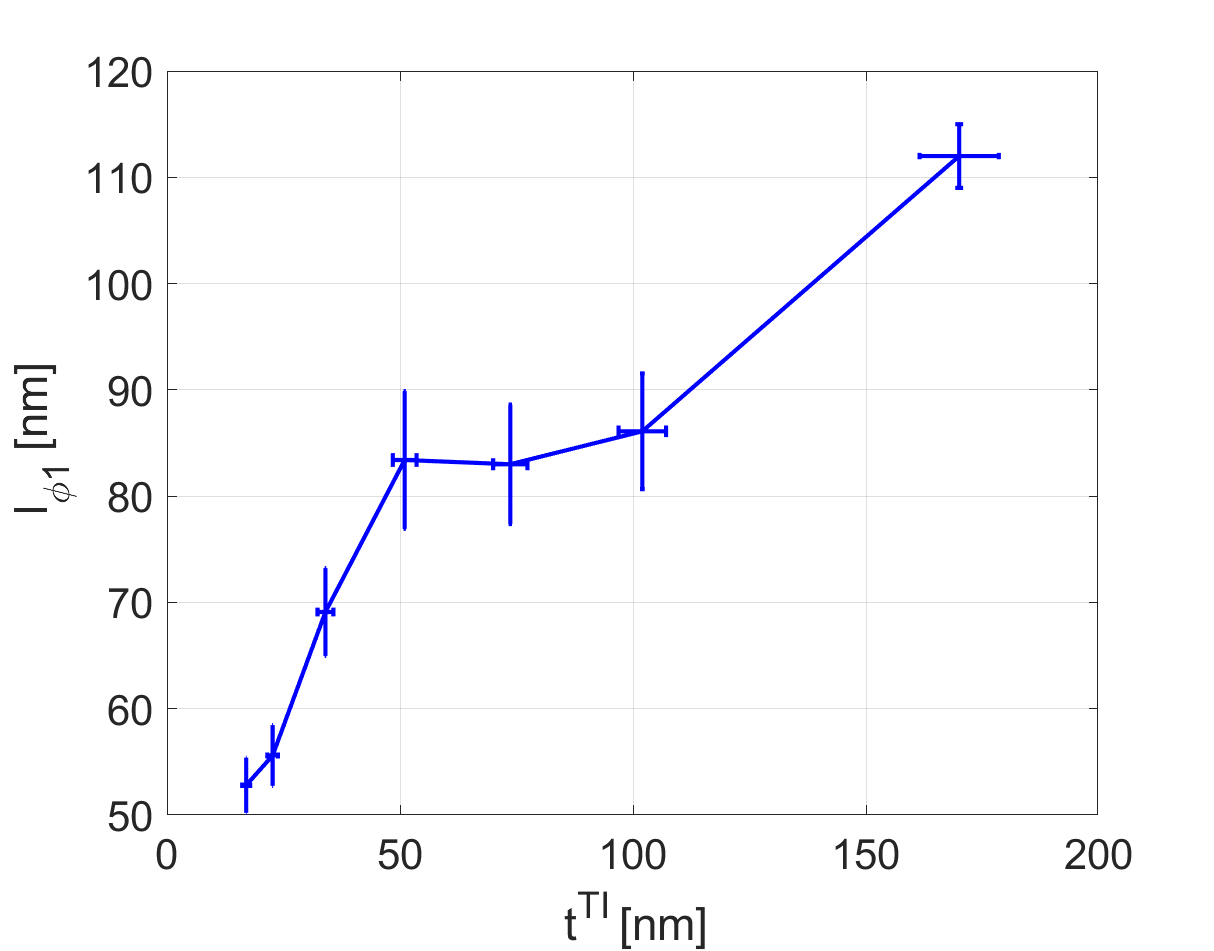}

}

\subfloat[\label{fig:fitted gap}]{\includegraphics[width=8.6cm]{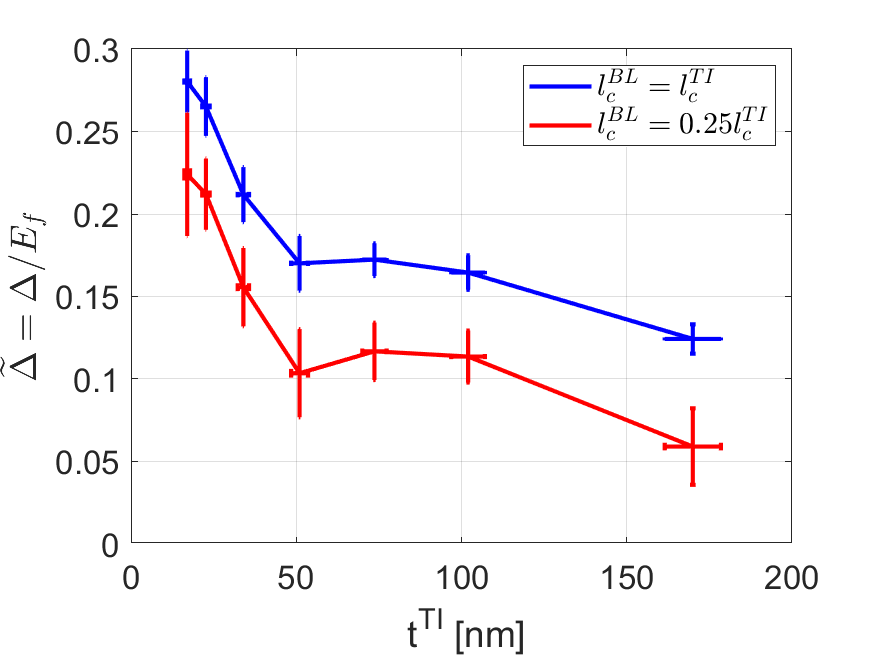}

}

\caption{(a) The bilayers' effective coherence length $l_{\phi1}$ vs TI layer thickness
$\left(t^{TI}\right)$, extracted by fitting the low field MC data
(Fig. \ref{fig:MC data with fit and a bare-TI bilayer comp}) using
Eq. \ref{eq:mHLN function}. (b) The bilayers' normalized
magnetically induced energy gap $\widetilde{\Delta}=\Delta/\epsilon_{F}$, obtained by fitting the low field MC data
assuming the bilayers' coherence length $l_{c}$ is the same
as in our bare TI films (blue), or fourth of this value (red). See main text for a discussion of this result. Error bars are from the fitting and
thickness uncertainty. Note the inflection point around 73 [nm] which we discuss in the main text. \label{fig:lphi1 and gap_tilde}}
\end{figure}

However, the two surfaces cannot couple if their coherence length is smaller than sample thickness\cite{Steinberg2011,Park2018}. Figure \ref{fig:fitted lphi1} shows that the coherence length is smaller than the thickness in the bilayers with 100-170 [nm] thick TI. Coupling between the surfaces should not occur in these samples and $\alpha$ should equal 1\cite{Steinberg2011,Park2018}, but this is not the case as mentioned above. We therefore have to conclude that there is one surface which dominates our MC data, and possibly smaller contributions from the other surface and the bulk. The surface states coherence length at the interface is generally smaller than at the exposed surface due to enhanced strain and defects at the interface\cite{He2011,Zhang2012}. In our bilayers there should be also stronger magnetic interactions at the interface which can further reduce the coherence length\cite{Lu2014,Lu2011}. We therefore conclude that the surface states at the exposed surface dominates our MC data. Consequently, the magnetically induced energy gap we are going to extract is a property of the exposed surface. 

This is a crucial point in our analysis. In appendix \ref{Detailed explanation appnd} we give another detailed justification for this conclusion. One important consequence is that we cannot explain the reduced bilayer MC as a result of magnetic scattering at the surface alone, because there are no magnetic particles at the exposed surface. Moreover, it allows us to use the MC data to study the thickness dependence of the magnetically induced energy gap, since the thickness should affect mainly the exposed surface states,
or the coupling of these states to the magnetic interface.

\textit{Normalized energy gap extraction:} In the m-HLN model the function $\alpha$ in Eq. \ref{eq:mHLN function}
depends solely on $\widetilde{\Delta}$\cite{Lu2014,Lu2011}, but
previous studies have shown that this is true only up to an experimental
prefactor, due to coupling between different transport channels and other interactions\cite{Kim2011,He2011,Zhang2012,Wang2016,Bansal2012}. The experimental prefactor masks 
the dependence of $\alpha$ on $\widetilde{\Delta}$, so we can regard $\alpha$ as an independent parameter. 
Consequently, there are two independent fit parameters in Eq. \ref{eq:mHLN function}:
$\alpha$ and the effective coherence length $l_{\phi1}$. Below we explain how we estimated the different contributions of 
the gap, the mean free path, and coherence length to the effective coherence length $l_{\phi1}$ (Eq. \ref{eq:lphi def}).

Figure \ref{fig:MC data with fit and a bare-TI bilayer comp} shows a strong effect of the FM 
layer on the TI surface states transport properties, as expected if the surface states are gapped\cite{Lu2014,Lu2011}.
The gap reduces the surface states coherence length (Eq. \ref{eq:lphi def}) and consequently reduces their MC (Eq. \ref{eq:mHLN function}).
Figure \ref{fig:fitted lphi1} shows the $l_{\phi1}$ values
for bilayers with various TI thicknesses and a constant 12 [nm] FM. 
The bilayers' $l_{\phi1}$ are of order 100 [nm], much shorter than in bare TI
films where we found $l_{c}$ of order 500 [nm]\cite{Lu2014,Lu2011}.
We therefore argue that the exposed surface states acquire a gap by a magnetic proximity effect,
and that this gap dictates the bilayer short effective coherence length, $l_{\phi1}$ (see Eq. \ref{eq:lphi def}).
Below we further justify these conclusions by showing self-consistency with other parts of our data. In particular,
we show that the gap is proportional to the magnetization we extract from the anomalous Hall data. We also explain below and in appendix \ref{xray data appendix} why we claim that the reduced effective coherence length is due to magnetic effects and not due to structural differences between the bare TIs and the TI layers in the bilayers.

To extract the normalized energy gap $\widetilde{\Delta}$ quantitatively we assumed that $l_{c}$ 
in the bilayers is the same as $l_{c}$ in the bare TIs.
Under this assumption, we used the full m-HLN equation to fit the MC data
and extract $\widetilde{\Delta}$ (see appendix \ref{Modified-HLN-formula} for details). Figure \ref{fig:fitted gap} shows $\widetilde{\Delta}$ of the bilayers as a function of TI thickness. 
The values we find correspond to energy gaps between $\sim0.28\epsilon_{F}$ and $\sim0.12\epsilon_{F}$. We also compare between the results we got by assuming that $l_{c}$ values in the bilayers equal $l_{c}$ of the bare TIs (blue), and the results we got for fourth of this value (red). As figure \ref{fig:fitted gap} shows, our assumption is robust and the results remain almost the same, as long as $l_{c}$ is at least 20\% of $l_{c}$ in the bare TIs. Our results show that the exposed surface states are gapped even in the thickest bilayer with a 170 [nm] TI layer, suggesting a coupling between the magnetic interface and the exposed surface states. As we will show below, this coupling highly depends on the TI bulk carrier density and thickness.

Figure \ref{fig:fitted gap} shows that $\widetilde{\Delta}$ has an inflection point as a function of thickness between 50 [nm] and 100 [nm]. Below we show that this corresponds to a minimum in the bulk magnetization and carrier density. To study this point, and since the magnetic coupling and $\widetilde{\Delta}$ could depend on the TI carrier density\cite{Li2015,Bruno1992,Kim2011,He2011,Wang2016,Bansal2012} and magnetization\cite{Lu2014,Lu2011}, we discuss now these properties.

\begin{figure}
\begin{centering}
\includegraphics[width=8.6cm]{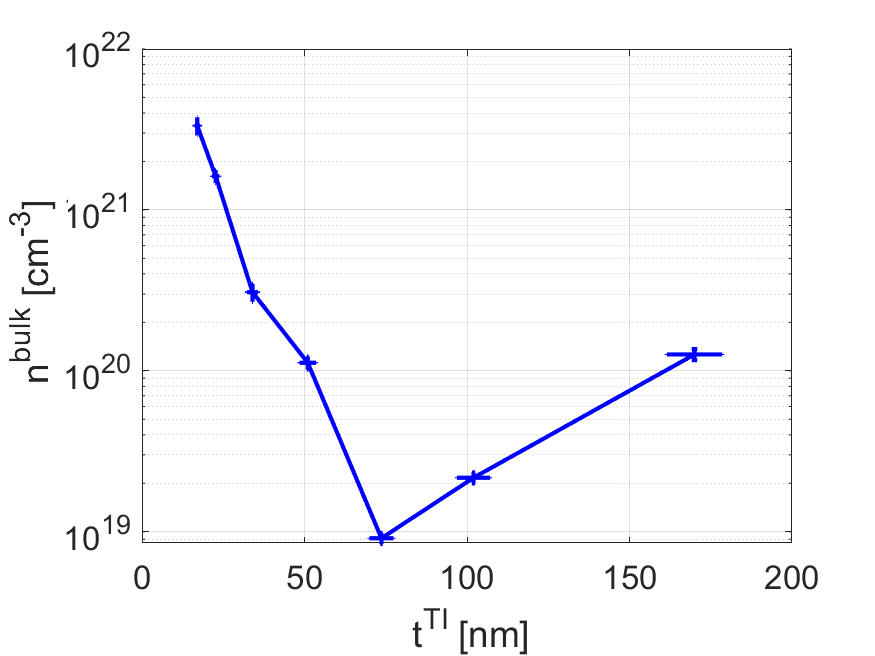} 
\par\end{centering}
\caption{The bilayer TI layer carrier density vs TI layer thickness calculated 
using Eq. \ref{eq:linear Hall equation}. The FM thickness is 12 [nm].
\label{fig:bilayer and TI hall coeff}}
\end{figure}

\textit{Thickness dependent bulk carrier density:} Figure \ref{fig:bilayer and TI hall coeff} shows the carrier density of the TI layer 
$\left(n^{bulk}\right)$ as a function of its thickness. Note that in the notation $\left(n^{bulk}\right)$ we mean $n$ of the TI bulk. 
Importantly, $n^{bulk}$ is minimal where the gap has its inflection point (Fig. \ref{fig:lphi1 and gap_tilde}).

We extracted $n^{bulk}$ from Hall measurements using the Drude model for conductors 
connected in parallel. We only needed to consider the FM layer and TI bulk 
transport channels because the high resistance of the surface states makes 
their contribution negligible \cite{Weyrich2016}. The Drude model gives: 
\begin{equation}
R_{H}^{tot}\equiv\dfrac{V_{xy}^{tot}}{BI}\sim\left(\dfrac{R_{xx}^{tot}}{R_{xx}^{FM}}\right)^{2}R_{H}^{FM}+\left(\dfrac{R_{xx}^{tot}}{R_{xx}^{bulk}}\right)^{2}R_{H}^{bulk}\label{eq:linear Hall equation}
\end{equation}

Where $R_{H}^{tot}$ and $R_{xx}^{tot}$ are the measured Hall coefficient and linear resistance of the bilayer. 
We used the approximation $R_{xx} = \left(nte\mu_{xx}\right)^{-1}$
with $t$ the thickness and $\mu_{xx}$ the mobility since in our samples $R_{xx}\gg R_{xy}$. 
Note that Eq. \ref{eq:linear Hall equation} holds only for the linear Hall terms without the anomalous
Hall terms which are a function of sample magnetization (see below). 
We extracted the linear terms from the high magnetic field data where the 
magnetization saturates.

To extract the TI bulk properties 
from Eq. \ref{eq:linear Hall equation} we need $R_{xx}^{FM}$ and $R_{H}^{FM}$. 
We could not measure them directly from a bare FM because the TI layer growth changes the FM
properties. We thus grew a bilayer with only 5 [nm] thick TI layer, where we assumed that 
the FM layer dominates the transport properties. This assumption is 
justified by the resistance of a bare 5 [nm] thick 
TI film being ten times larger than that of
a 12 [nm] thick FM film. Therefore, we used the measured 
properties of the BL with 5 [nm] TI to represent the properties of the FM layer in our analysis. 
For example, we took $R_{H\left(t^{TI}=5 \:\left[nm\right]\right)}^{tot}\approx R_{H}^{FM}$.

The carrier density non-monotonic behavior (Fig. \ref{fig:bilayer and TI hall coeff})
may result from a combination of different reasons. First, lattice mismatch creates strain and defects that decrease with film thickness. Second, the metal-semiconductor
interface leads to a band bending due to the Shottky barrier effect\cite{BartVanZeghbroeck2008}.
Third, $Se$ diffusion from the $Fe_{7}Se_{8}$ layer can reduce hole
concentration since $Se$ is an electron donor in $BST$ compounds\cite{Ren2011,Arakane2012,Neupane2012}.
Different thickness dependence of the various contributions to the
carrier density variation can lead to a non-monotonic carrier density
thickness dependence. However, to quantify these effects will require
a detailed chemical and structural study, which is beyond the scope of this paper.

\textit{Bulk magnetization:} We turn now to the magnetization. In a FM, the Hall
resistivity includes an anomalous Hall effect (AHE) term that is proportional to perpendicular magnetization\cite{Shiomi2013,Nagaosa2010}:
\begin{equation}
\rho_{xy}\equiv\dfrac{V_{xy}}{I_{xx}}t=R_{H}^{tot}tB_{z}+\alpha^{tot}M_{z}^{tot}\label{eq:full Hall equation}
\end{equation}

With $B_{z}$ and $M_{z}^{tot}$ being the out of plane component of magnetic field and bilayer magnetization, respectively.
According to the theory of Luttinger and Berger, $\alpha^{tot}=C*\rho_{xx}^{\beta}$
with $\beta\sim2$ and $C$ a constant. For simplicity we take $\beta=2$. 

\begin{figure}
\subfloat[\label{fig:AHE sym}]{\includegraphics[width=8.6cm]{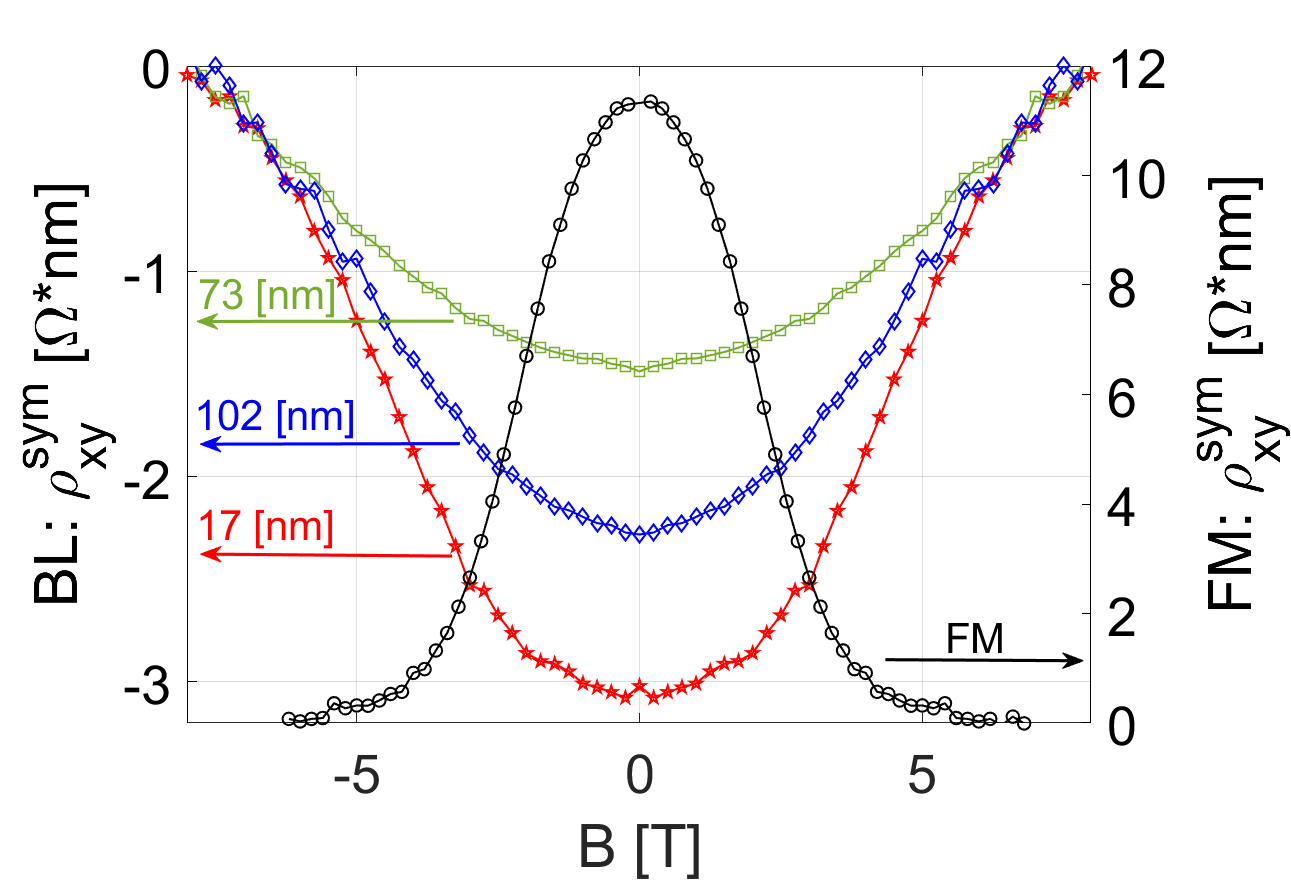}

}

\subfloat[\label{fig:rem_mag vs t}]{\includegraphics[width=8.6cm]{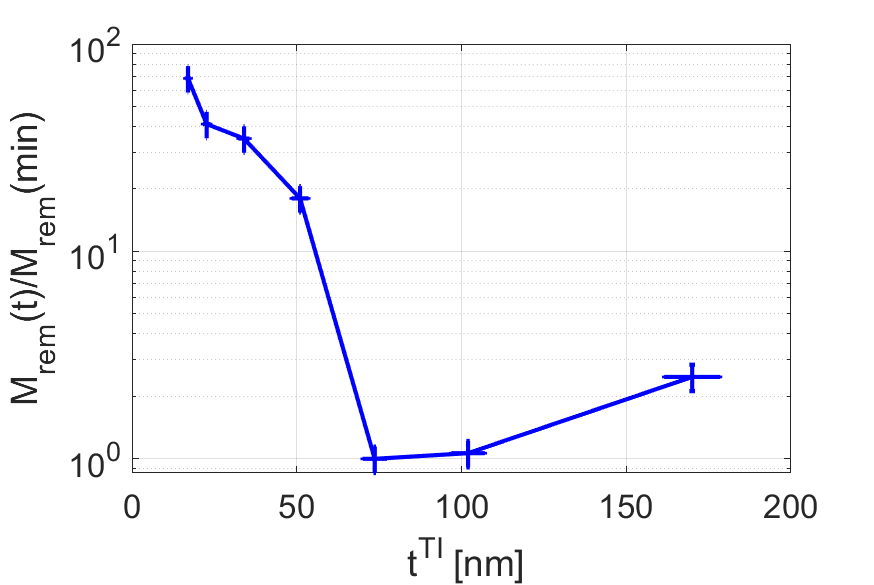}

}

\caption{(a) The bilayer (total) Hall resistivity symmetric part, which is
the anomalous Hall effect, for bare $Fe_{7}Se_{8}$ (black, right
y axis) and bilayers with different TI thickness (left y axis):
17 [nm] (red), 73 [nm] (green), 102 [nm] (blue). Note the opposite
sign of the $Fe_{7}Se_{8}$ and bilayer signals. The displayed data is
for positive to negative magnetic field direction in the hysteresis
curve. (b) The bilayer remnant magnetization vs thickness normalized by the minimal value 
to cancel unknown constants. As explained in main text, the signal
arises mainly from the magnetization that the FM induces in the TI
layer.}
\end{figure}

To differentiate between the ordinary and anomalous terms we take the symmetric part 
with respect to the magnetic field: $\rho_{xy}^{sym}\equiv\left(\rho_{xy}\left(B\right)+\rho_{xy}\left(-B\right)\right)/2=C*\left(\rho_{xx}^{2}M_{z}\right)^{sym}$.
Figure \ref{fig:AHE sym} shows $\rho_{xy}^{sym}$ for a bare 12 [nm]
thick $Fe_{7}Se_{8}$ film (black, right axis) and three bilayers (colored,
left axis). 

Intriguingly, the bilayers have a negative AHE term while in the bare Fe$_{7}$Se$_{8}$
the sign is positive. The bilayer curves FWHM is 
wider, namely their coercive fields are higher than that of the bare Fe$_{7}$Se$_{8}$.

Previous works have shown negative AHE terms in TI/FM-insulator bilayers,
where the AHE arises from the TI layer \cite{Lang2014,Tang2017,Yang2019}.
They have also shown that the FM magnetization penetrates the TI layer
with a flipped direction, explaining the negative AHE\cite{Lang2014}.
Bulk mediated magnetic exchange coupling is also diamagnetic in $Bi$
compounds\cite{Li2015}. We therefore conclude that the TI layers have
the dominant contribution to the AHE of the bilayers. 
The AHE in the TI is indicative of time reversal
symmetry breaking and gapped surface states\cite{Yang2013,Tang2017},
as we argued by analyzing the effective coherence length (Fig. \ref{fig:fitted lphi1}).

We are interested in the zero field surface states energy gap so we will
focus now on the remnant magnetization, defined as the magnetization at zero external magnetic field.
Using Eq. \ref{eq:linear Hall equation} and Eq. \ref{eq:full Hall equation} we can write (appendix \ref{Mrem appendix}):
\begin{equation}
M_{rem}\equiv M_{z}\left(0\right)\sim C^{-1}\left(\dfrac{R_{xy}^{tot}\left(0\right)}{t^{TI}\left(R_{xx}^{tot}\left(0\right)\right)^{2}}\right)\label{eq:Magnetization final}
\end{equation}
Figure \ref{fig:rem_mag vs t} shows $M_{rem}$ vs TI thickness normalized
by its minimal value. A non-zero value of $M_{rem}$ leads to a gap
in the surface states. Notably, the remnant magnetization $M_{rem}$ and carrier
density $n^{bulk}$ (Fig. \ref{fig:bilayer and TI hall coeff}) share a minimum just
where the normalized magnetic gap has its inflection (Fig. \ref{fig:lphi1 and gap_tilde}).

\textit{Analyzing the gap using bulk properties:} So far we only discussed the normalized gap, $\widetilde{\Delta}\equiv\dfrac{\Delta}{\epsilon_{F}}$,
which we obtained from the m-HLN fit. To estimate the actual gap values
we first need to estimate $\epsilon_{F}$. Using the parabolic band approximation, we
can write:
\begin{equation}
\epsilon_{F}=\dfrac{\hbar^{2}}{2m^{*}}\left(3\pi^{2}n^{TI}\right)^{\nicefrac{2}{3}}\label{eq:parabolic approximation}
\end{equation}
Here, $m^{*}$ is the effective electron mass
and $n^{TI}$ is the carrier density of the TI layer (Fig. \ref{fig:bilayer and TI hall coeff}).

\begin{figure}
\subfloat[\label{fig:extracted gap}]{\includegraphics[width=8.6cm]{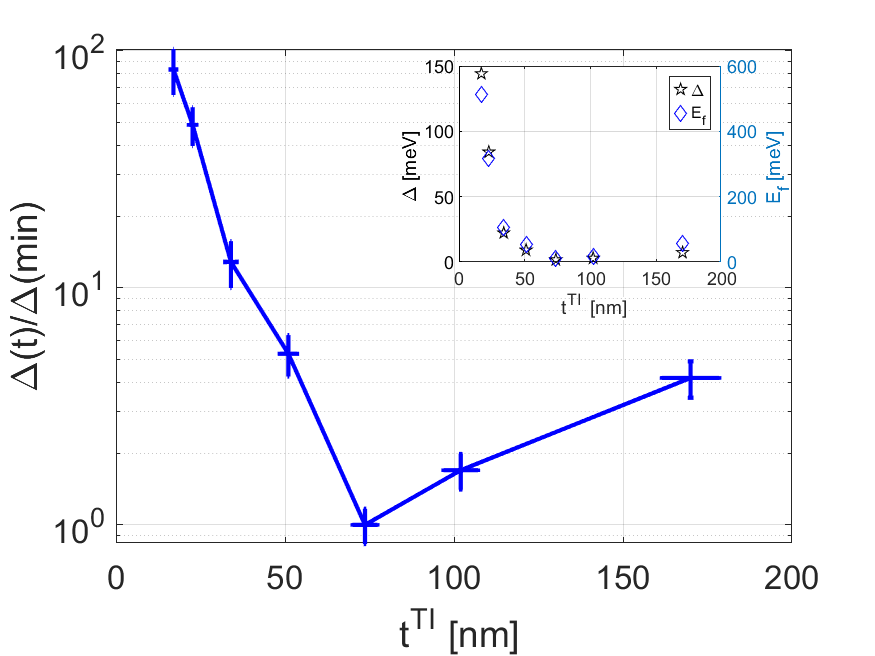}

}

\subfloat[\label{fig:M_and_gap_comp}]{\includegraphics[width=8.6cm]{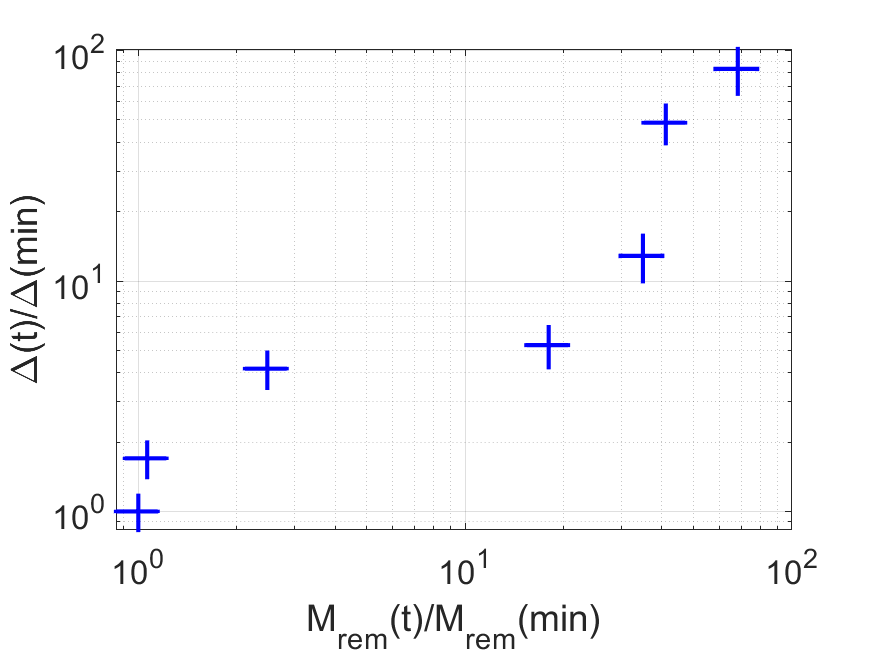}

}

\caption{(a) The magnetic gap $\Delta$ vs $t^{TI}$ normalized by its
minimal value, showing a similar thickness dependence as the magnetization
and carrier density (Figures \ref{fig:bilayer and TI hall coeff} and \ref{fig:rem_mag vs t}). We calculated $\Delta$ assuming a parabolic band. Inset: The Fermi energies and $\Delta$ values vs $t^{TI}$. These are only rough estimates since there are some parameters we do not know such as the shift between the Dirac point and the valence band. However, different values of those parameters did not change the thickness dependence.
(b) The magnetic gap $\Delta$ vs the remnant bulk magnetization $\left(M_{rem}\right)$.
We normalized the values by the minimal value to cancel unknown constants.
Importantly, the gap depends monotonically on the bulk magnetization.}
\end{figure}

Figure \ref{fig:extracted gap} shows the magnetic gap $\Delta$
as a function of the TI layer thickness. We normalized the values
by the minimal value to cancel the mass prefactor. 
Figure \ref{fig:M_and_gap_comp} shows
a clear correlation between the bulk magnetization and the energy gap. 
We note that this correlation verifies our assumption that the clear difference between 
the magneto-conductivity of the bilayers and bare TI films
(Fig. \ref{fig:MC data with fit and a bare-TI bilayer comp}) is due to a gap opening in the surface states. 
The gap must be opened since the sample magnetization breaks time reversal symmetry\cite{Bernevig2013,Hasan2010}.

Previous works found that the bulk states effective electron mass
in $Bi_{2-x}Sb_{x}Te_{3}$ compounds is about one electron mass\cite{Madavali2019,Whitney2017}.
Inserting $m^{*}=m_{e}$ in Eq. \ref{eq:parabolic approximation}
we get $\epsilon_{F}$ values of ~10-400 [meV] and corresponding magnetic
gap values of ~1.5-120 [meV], depending on TI thickness. These gap values
are similar to gaps reported in other TI/FM systems\cite{Okuyama2019,Yang2019,Zheng2016}. 
However, the parabolic band approximation does not fully describe the band structure of $Bi_{2-x}Sb_{x}Te_{3}$,
so the real gap values may differ from our estimation.

The results indicate that there is a gap even for the thickest TI layer of 170 [nm].
A natural length scale for exchange interactions is\cite{Bruno1992}
\begin{equation}
l_{int}=\dfrac{\hbar v_{F}}{2\pi k_{B}T}\label{eq:lintarticle}
\end{equation}
Using the parabolic band approximation with $m^{*}=m_{e}$ and our
carrier densities we get length scales of ~45-300 [nm]. For example, $l_{int}$ is about 100 [nm] in
our 170 [nm] sample. These length scales are of the same order of magnitude as the 
samples thickness which can explain how we found proximity effect is such thick samples.

The minimum in the energy gap that we find for a specific thickness may seem surprising. 
One would expect the interactions between the exposed surface states 
and the magnetic interface to decay monotonically
with the TI thickness. However, the interaction length scale $l_{int}$
depends on $v_{F}$ which is a function of the carrier density. This
can result in a non-monotonic thickness dependence of the exchange
interaction, because the carrier density dependence on the thickness is non monotonic in 
our bilayers (Fig. \ref{fig:bilayer and TI hall coeff}).

In our bilayers we grew the TI layers on the FM layer, while we grew the bare TIs directly on the $Al_{2}O_{3}$ substrate. Therefore one might assume that the reduced surface states effective coherence length (figures \ref{fig:MC data with fit and a bare-TI bilayer comp} and \ref{fig:lphi1 and gap_tilde}) is due to stronger strains and smaller grain size, and not due to magnetic proximity effect as we suggest here. Indeed, using X-ray data we show in appendix \ref{xray data appendix} that there is slightly higher strain in the TI layers of the bilayers compared to the bare TIs. However, we show that this is a tensile in-plane strain, which is known to enhance surface properties and therefore cannot reduce the coherence length in the bilayers \cite{Liu2014}. We also show in appendix \ref{xray data appendix} that the TI grain size is about ~45 [nm] both in the bare TIs and the bilayers, confirming that grain size difference is not responsible for our results. Moreover, in appendix \ref{Fe doping appendix} we explain why we could neglect possible effects of Fe doping of the TI layer from the underlining FM layer in our analysis.

Previous publications have shown that ferromagnetism penetrates only
~2 [nm] into the TI\cite{Katmis2016,Li2017}. This might seem
in contradiction with our results. However, in these publications the films have very low carrier concentration of $n^{bulk}\lesssim10^{19} \:\left[cm^{-3}\right]$, and the measurements were done at a slightly higher temperature where equation \ref{eq:lintarticle} indeed predicts a small magnetization penetration length of $l_{int}\lesssim10 \:\left[nm\right]$. More importantly, these publications present induction of ferromagnetism into the TI, while RKKY-like interactions are antiferromagnetic. It might be that only a small portion of our TI layers is ferromagnetically magnetized and gives the AHE signal. However, the exposed TI surface can still interact with this ferromagnetic part via bulk states mediated long ranged RKKY-like antiferromagnetic interactions\cite{Li2015,Bruno1992}. Such interactions break the time reversal symmetry of the exposed surface states \cite{He2017} and open a gap.

\section{Summary}

In this study we measured the electrical transport properties of doped-TI/FM-metal
bilayers, made of 17-170 [nm] $Bi_{0.8}Sb_{1.2}Te_{3}$ on 12 [nm] $Fe_{7}Se_{8}$.
By fitting the low field MC data to the m-HLN model we demonstrated its validity in doped-TI/FM-metal interfaces.
As a fit parameter we extracted the exposed surface states effective coherence
length $l_{\phi1}$, which is a geometric average of the coherence length 
and a term arising from a magnetically induced gap. 

We compared the MC and $l_{\phi1}$ between our bilayers and bare TIs and showed that $l_{\phi1}$ in the bilayers is much shorter.
We therefore argued that the gap term dominates $l_{\phi1}$ in the bilayers. 
Assuming that the ordinary coherence length $l_{c}$ of the surface states in the bilayers is not changed by more than a factor of five compared to its value in the bare TIs, we extracted 
the normalized gap $\widetilde{\Delta}=\Delta/\epsilon_{F}$.
We found that the normalized energy gap of the 
exposed surface states $\widetilde{\Delta}$
remains finite even for a 170 [nm] thick TI layer. This indicates
a coupling between the magnetic interface and the exposed
surface states which is mediated by the bulk states. We conclude that
high bulk carrier densities of $n_{bulk}\gtrsim10^{20} \:\left[cm^{-3}\right]$ are necessary 
for the coupling in such thick layers.

To justify our analysis, we first showed a strong correlation between
the carrier density the magnetization and $\widetilde{\Delta}$.
Second, we showed that the bilayer anomalous Hall signal
is dominated by the magnetic interactions inside the TI layer.
Such magnetic interactions break time reversal
symmetry in the TI and open a gap
in the surface states. Lastly, using a simple model for the electronic dispersion we showed that for thin bilayers (where data by other groups exists) 
the gap values we found are comparable to values found in other TI/FM systems\cite{Okuyama2019,Yang2019,Zheng2016}.

More studies with different
FM materials and TI thicknesses and carrier densities, as well as direct
energy band measurements, are needed to fully understand the physical mechanism behind the carrier density and thickness dependent magnetic proximity effect we found in this study.

\section{Methods}
We grew our bilayers on polished $Al_{2}O_{3}(0001)$ substrates of a $1 \left[cm^{2}\right]$
area and 0.1 [cm] thickness. The average substrate RMS roughness was
 below 0.5 [nm]. Before each bilayer growth we used a mask during the deposition
to pattern the bilayers into a VDP compatible geometry. After annealing
the substrate at 330 [C] for an hour, we had used a
homemade pulsed laser deposition (PLD) device to grow the bilayer. The
base pressure of the deposition chamber was $\sim1*10^{-7} \:\left[Torr\right]$. The substrate temperature was 280 [C] for both
$Fe_{7}Se_{8}$ and $B_{0.8}Sb_{1.2}Te_{3}$ layers. In the PLD process
we used a 355 [nm] Nd:YAG laser with a ~1 [Hz] repetition rate
for the $Fe_{7}Se_{8}$ and ~3 [Hz] for the $B_{0.8}Sb_{1.2}Te_{3}$.
The pulse energy flux was $0.33 \:\left[\frac{J}{cm^{2}}\right]$ and $0.2 \:\left[\frac{J}{cm^{2}}\right]$
respectively. The $Fe_{7}Se_{8}$ target was a pressed pellet of powders
of $Fe$ and $Se$ with a ratio of $3:10$. The $B_{0.8}Sb_{1.2}Te_{3}$
target was a pressed pellet made of grounded $B_{0.7}Sb_{1.3}Te_{3}$
single crystals.

We characterized the bilayers orientation and phase using a Bruker D8$_{eco}$
X-ray diffractometer, and found the characteristic c-axis peaks of
the $Fe_{7}Se_{8}$ and $B_{0.8}Sb_{1.2}Te_{3}$ phases. We measured
thickness and roughness using a Veeco Wyko NT1100 profilometer. The
characteristic bilayer RMS roughness was  below 2 [nm]. To verify the bilayers stoichiometry
and homogeneity we used a Quanta SEM-EDS. We did not find any
detectable chemical inhomogeneity up to the device resolution.
To further analyze the thickness dependent chemical composition and
the quality of the interface we used TOF-SIMS. We
discussed the results in the main text.

For electrical transport measurements we used an Oxford Teslatron
and homemade electronics. We made the contacts from silver paint in
a 4-contacts Van der Pauw configuration (VDP). The bilayer's area
was ~1*1 [mm], and the thicknesses were 5-170 [nm]. The contacts
area was less than $5\%$ of the sample size and reciprocity theorem
was valid within $1\%$, which hold with VDP measurement requirements.
We conduct all magneto-transport measurements with DC source current
of 0.75 [mA] at 1.5 [K] with temperature fluctuations of $\Delta T\le5 \:\left[mK\right]$.

\section{Data availability}
The data that support our findings are available from the corresponding
author upon reasonable request.

\section{Acknowledgments}
We thank B. Shapiro and E. Akkerman for useful discussions. 
This work was supported by the Israel Science Foundation grant no. 320/17. 
Y. Jarach was supported by the Israel Ministry of Energy under the program of BSc to PhD students scholarships in the field of energy.

\appendix

\section{The Modified HLN formula for Surface States \label{Modified-HLN-formula}}

For the massive Dirac cone dispersion, the magnetoconductivity can
be fitted by a modified HLN formula\cite{Lu2014,Lu2011} with an experimental
global pre-factor\cite{Kim2011,He2011,Zhang2012,Wang2016,Bansal2012}: 
\begin{equation}
\sigma{\left(B\right)}-\sigma(0)=A\sum_{i=1,2}\frac{\alpha_{i}e^{2}}{\pi h}\left[\psi\left(\frac{1}{2}+\frac{l_{B}^{2}}{l_{\phi i}^{2}}\right)-ln\left(\frac{l_{B}^{2}}{l_{\phi i}^{2}}\right)\right]\label{eq:full m-HLN formula}
\end{equation}
Where A is the experimental prefactor, $l_{B}$ is the magnetic length,
$l_{B}=\sqrt{\frac{\hbar}{4e\left|B\right|}}\sim\frac{13}{\sqrt{\left|B\right|}}\left[\frac{nm}{\sqrt{T}}\right]$
and $l_{\phi i}^{-2}=\frac{l_{c}^{2}l_{i}^{2}}{l_{c}^{2}+l_{i}^{2}}$,
$l_{c}$ is the surface states' phase coherence length, and $\psi=\partial_{x}ln\left(\Gamma_{\left(x\right)}\right)$
(the digamma function) . For $\cos\left(\theta\right)=\frac{\Delta_{gap}}{2E_{Fermi}}\equiv\frac{\widetilde{\Delta}}{2}$,
$l_{e}$ the mean free path and $\left(a,b\right)=\left(\cos\left(\frac{\theta}{2}\right),\sin\left(\frac{\theta}{2}\right)\right)$,
the $l_{i}$ and $\alpha_{i}$ parameters are given by\cite{Lu2011}:
\begin{eqnarray}
l_{1}^{2} & = & l_{e}^{2}\left(a^{4}+b^{4}\right)^{2}\left[a^{2}b^{2}\left(a^{2}-b^{2}\right)^{2}\right]^{-1}\\
\alpha_{1} & = & -a^{4}b^{4}\left[\left(a^{4}+b^{4}-a^{2}b^{2}\right)\left(a^{4}+b^{4}\right)\right]^{-1}\\
l_{2}^{2} & = & l_{e}^{2}a^{4}\left(a^{4}+b^{4}-a^{2}b^{2}\right)\left[b^{4}\left(a^{2}-b^{2}\right)^{2}\right]^{-1}\\
\alpha_{2} & = & \left(a^{4}+b^{4}\right)\left(a^{2}-b^{2}\right)^{2}\left[2\left(a^{4}+b^{4}-a^{2}b^{2}\right)^{2}\right]^{-1}
\end{eqnarray}
For pure weak antilocalization, namely without magnetic interactions\cite{Lu2014,Lu2011},
$\left(\alpha_{1},\alpha_{2}\right)=\left(-0.5,0\right)$ . By fitting
the magnetoconductivity data with this formula the normalized magnetically
induced energy gap, $\widetilde{\Delta}$, can be extracted. This
formula assumes $l_{e}\ll l_{c}$ and $l_{e}\ll l_{B}$\cite{Lu2014,Lu2011},
which limits its validity only for $B\lesssim1T$ in topological insulators
where $l_{e}\sim10-100 \:\left[nm\right]$\cite{Lu2011}. It also assumes that the
phase coherence length includes all the information about disorder
and scattering from magnetic impurities\cite{Lu2011}.

The full model have four fit parameters- A, $\widetilde{\Delta}$,
$l_{e}$ and $l_{c}$. In the low magnetically induced energy gap
limit $\alpha_{2}$ is smaller than $\alpha_{1}$ and the MC is mainly
dominated by the $\alpha_{1}$ and $l_{\phi1}$ terms. In this case
the experimental prefactor almost masks the $\alpha_{1}$ dependence on
$\widetilde{\Delta}$ since the prefactor also depends on sample properties. In addition, $l_{\phi1}$ depends on parameters
$\widetilde{\Delta}$, $l_{e}$ and $l_{c}$. Hence, in the low
gap limit there are two dominant effective fit parameters, A$\alpha_{1}$
and $l_{\phi1}$, and one cannot get a trustworthy MC fit using all
four parameters. To simplify notations we define $A\alpha_{1}\equiv\alpha$ when we discuss this limit
in the article (Eq. 1).

In this article we expect energy gaps in the low gap regime, so we
begin with analysis of this limit (see the discussion of Eq. 1 and its results in the main text). 
This is self-consistent with our results. However, to disentangle the gap term $\widetilde{\Delta}$
from the mean free path $l_{e}$ we need the full model (Eq. \ref{eq:full m-HLN formula}). To do this despite
the low gap limit fitting problems just discussed, 
we first show that our results really indicate that such
a gap indeed exists. Next, we show that the bilayer fitted $l_{\phi1}$ values
are about five times smaller than the real coherence length $l_{c}$ in bare
TIs. This allows us to assume that in the bilayer, the real coherence length contribution to 
$l_{\phi1}$ is small.
Therefore, in the bilayers we did not use $l_{c}$ as a fit parameter but 
substitute its values with the values obtained for bare TI films. This process is valid unless the bilayer 
$l_{c}$ is about five times smaller than in the bare TI films. After concluding that 
a gap exists, and substituting $l_{c}$, we only used the full model to separate the contributions of
$l_{e}$ and $\widetilde{\Delta}$ to $l_{\phi1}$ and $l_{\phi2}$.

\section{Calculation of the remnant magnetization\label{Mrem appendix}}

In a system where only the ordinary Hall effect takes place the transverse resistance is $R_{xy}=R_{H}B_{z}$, with $R_{H}\equiv1/\left(nte\right)$;
$B_{z}$ the out-of-plane magnetic field; $n$ the 3D carrier density; 
$t$ the thickness; and $e$ the electron charge\cite{Shiomi2013,Nagaosa2010}. 
As explained in the main text (see Eq. 3 in the main text), we can get the total transverse (Hall) resistance of our bilayers by using the Drude model for parallel connection of the FM layer and the TI bulk:
\begin{equation}
R_{xy}^{tot}=\left(\dfrac{R_{xx}^{tot}}{R_{xx}^{FM}}\right)^{2}R_{xy}^{FM}+\left(\dfrac{R_{xx}^{tot}}{R_{xx}^{bulk}}\right)^{2}R_{xy}^{bulk}\label{eq:appndix linear Hall}
\end{equation}
where $R_{xy}^{tot}$ and $R_{xx}^{tot}$ are the values we got from measurements of the transverse and parallel resistance of a bilayer, respectively. 
The TI surface states are negligible here due to their high parallel resistance $R_{xx}$.

In a general system with the anomalous Hall effect, the transverse resistance takes the form\cite{Shiomi2013,Nagaosa2010}:
\begin{equation}
R_{xy}=R_{H}B+\dfrac{\alpha M_{z}}{t}\label{Rxy tot appnd}
\end{equation}
$M_{z}$ is the out-of-plane magnetization component. In the main text we discuss $\alpha$ and explain that according to anomalous Hall effect theories we can approximate $\alpha\sim C\rho_{xx}^{2}$ with C a constant\cite{Shiomi2013,Nagaosa2010}. Since $\rho_{xx}=R_{xx}t$, we can write $\alpha\sim C\left(R_{xx}t\right)^{2}$.

In the main text we conclude that the TI layer has the 
largest contribution to the AHE of the bilayers.
Therefore, we neglect the anomalous term of the FM layer. Now, in Eq. \ref{eq:appndix linear Hall} we can replace $R_{xy}^{FM}$ with $R_{H}B_{z}$ (ordinary Hall term), and $R_{xy}^{bulk}$ with Eq. \ref{Rxy tot appnd}. We therefore get:
\begin{equation}
R_{xy}^{tot}=\left(\dfrac{R_{xx}^{tot}}{R_{xx}^{FM}}\right)^{2}R_{H}^{FM}B_{z}+\left(\dfrac{R_{xx}^{tot}}{R_{xx}^{bulk}}\right)^{2}\left(R_{H}B_{z}+\dfrac{(\alpha M_{z})^{bulk}}{t^{TI}}\right)\label{eq:appndix total Hall after subs}
\end{equation}
We are interested in the remnant magnetization, namely the magnetization at zero external magnetic field. Substituting $B=0$ we get:
\begin{equation}
R_{xy}^{tot}\left(0\right)=\left(\dfrac{R_{xx}^{tot}\left(0\right)}{R_{xx}^{bulk}\left(0\right)}\right)^{2}\left(\dfrac{\alpha^{bulk}\left(0\right) M_{z}^{bulk}\left(0\right)}{t^{TI}}\right)
\end{equation}
Since the magnetization term comes from the TI, then $\alpha\sim C\left(R_{xx}^{bulk}t^{TI}\right)^{2}$ so finally we get:
\begin{equation}
R_{xy}^{tot}\left(0\right)\sim C\left(R_{xx}^{tot}\right)^{2}t^{TI}M^{bulk}_{z}
\end{equation}
Now we can extract the remnant magnetization:
\begin{equation}
M_{rem}\equiv M_{z}\left(0\right)\sim C^{-1}\left(\dfrac{R_{xy}^{tot}\left(0\right)}{t^{TI}\left(R_{xx}^{tot}\left(0\right)\right)^{2}}\right)
\end{equation}

\section{A simple model for the remnant magnetization \label{Rem mag appnd}}

\begin{figure}
\subfloat[\label{fig:lint}]{\includegraphics[width=8.6cm]{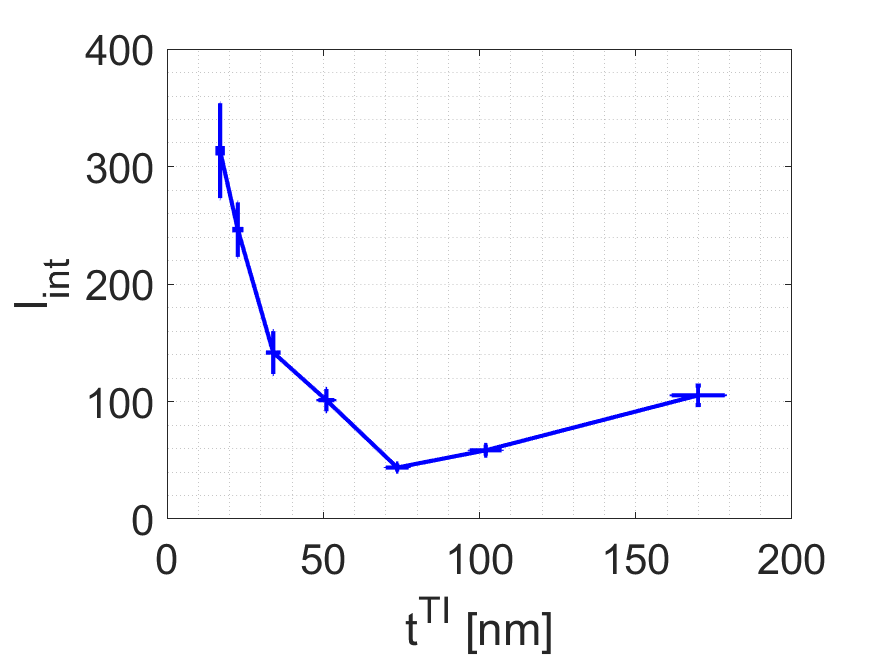}

}

\subfloat[\label{fig:Mav}]{{\includegraphics[width=8.6cm]{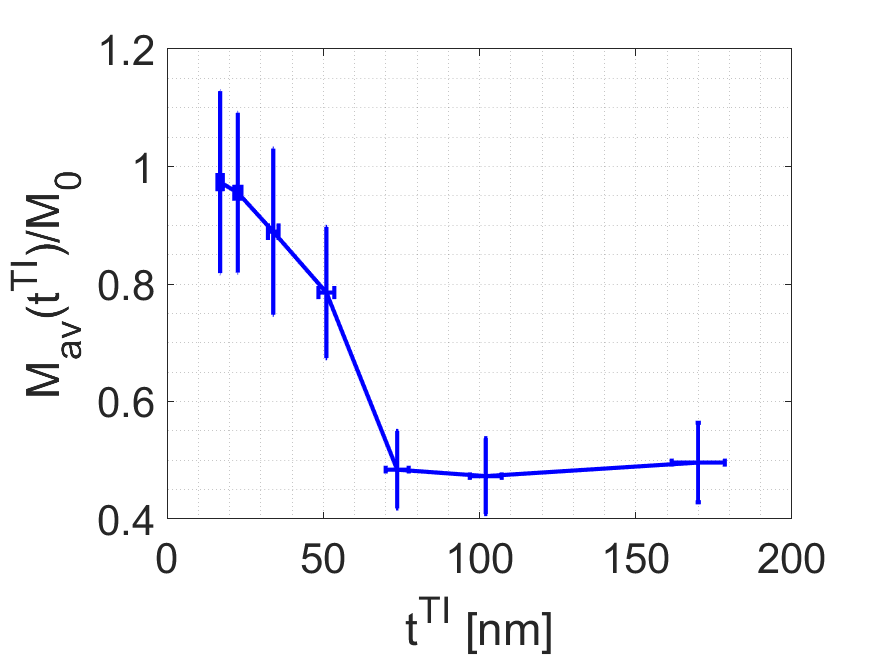}}

}

\caption{(a) The magnetic interaction length $l_{int}$ vs the TI layer thickness in our bilayers. See the text for $l_{int}$ definition. (b) The averaged magnetization $M_{av}$ of our model as defined in the main text vs the TI layer thickness.}
\end{figure}

To get some insights into the non-monotonic thickness dependence of the bilayers remnant magnetization
(and as a result, of the energy gap), we use a simple exponential decay model:
\begin{equation}
M(t)=M_{0}exp\left(-\dfrac{t}{l_{int}}\right)\label{eq:Mmodel}
\end{equation}
Here $M(t)$ is the sample magnetization and $t$
is the distance from the FM layer. $M_{0}$ is the magnetization at the FM-TI interface. 
We assume that the FM dictates $M_{0}$ such that it is sample independent. 
The magnetic interaction length $l_{int}$ is:
\begin{equation}
l_{int}=\dfrac{\hbar v_{F}}{2\pi k_{B}T}\label{eq:lint}
\end{equation}
where $v_{F}$ is the Fermi velocity and $T$ is the temperature. 
We can calculate $l_{int}$ under the parabolic band approximation 
using Eq. \ref{eq:lint} and the TI bulk carrier densities (Fig. 4 in the main text).
Fig. \ref{fig:lint} shows $l_{int}$ of the bilayers vs the TI layer thickness.
Up to our thickest TI layer (170 [nm] thick) the magnetic interaction length 
remains of the same order as the TI thickness. 
This explains how the gap is finite even in such thick samples.

As Eq. \ref{eq:Mmodel} presents, the magnetization induced into the TI decays with the distance from the FM-TI interface. The remnant magnetization we got from the anomalous Hall effect measurements (Fig. 5b in the main text) is though an averaged value over the TI thickness. We define the averaged magnetization in our model as:
\begin{equation}
M_{av}(t^{TI})=\dfrac{\int_{0}^{t^{TI}}M{\left(z\right)}dz}{\int_{0}^{t}dz}=M_{0}\dfrac{l_{int}}{t^{TI}}\left(1-exp\left(-\dfrac{t^{TI}}{l_{int}}\right)\right)\label{eq:modeled magnetization}
\end{equation}
Fig. \ref{fig:Mav} shows $M_{av}\left(t^{TI}\right)$ for our bilayers against the TI layer thickness in units of $M_{0}$. We get that $M_{av}\left(t^{TI}=170 \:\left[nm\right]\right)>M_{av}\left(t^{TI}=73 \:\left[nm\right]\right)$. As can be seen from Eq. \ref{eq:modeled magnetization} and Eq. \ref{eq:lint}, this inequality and the non-monotonic behavior of the magnetization 
(figures \ref{fig:rem_mag vs t} and \ref{fig:Mav}) 
result from the dependence of $l_{int}$ on 
the BL thickness, which in turn results from the carrier density dependence 
on the BL thickness (Fig. 4 in the main text).

\section{Role of Iron doping in the TI layer\label{Fe doping appendix}}

Tof-SIMS measurements put an upper limit of about $1\%$ on Fe dopant
concentration in the TI layers. This is far below the $\sim6\%$ Fe
doping needed to open a gap in the surface states dispersion\cite{Ribak2014}. In
addition, our out-of-plane coercive fields are almost an order of
magnitude larger than in Fe-doped TIs\cite{Ji2012}. Last, Fe dopant
in BSTS compounds are known as electron donors\cite{Zhao2019,Sugama2001}.
Since our samples are p-type, Fe doping reduces the charge density.
If the Fe dopants are dominant in the bilayer magnetic properties,
the magnetization should be maximal where the charge density is minimal
(maximal Fe density). This is opposite to our results where the charge
density, magnetization, and magnetization induced gap share the same
minimum. We therefore conclude that Fe dopants are not dominant in
the bilayers magnetization. We claim that the dominant mechanism is
interactions between the TI bulk, the upper TI surface, and the FM
layer, which the TI conductive bulk states mediate.

\section{Role of strain and grain size\label{xray data appendix}}

\begin{figure}
\subfloat[\label{fig:Xray}]{\includegraphics[width=8.6cm]{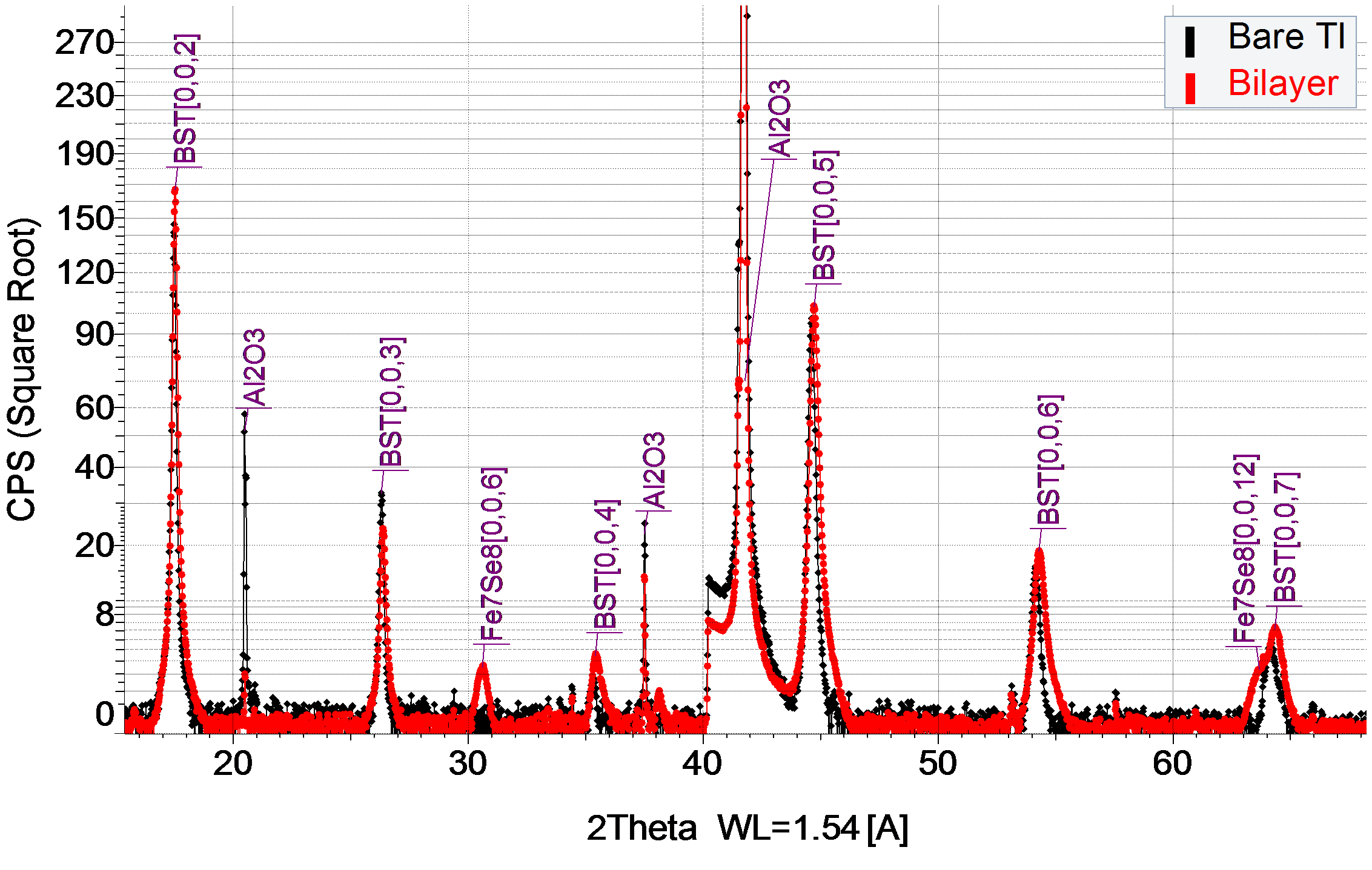}

}

\subfloat[\label{fig:Xrayzoom}]{{\includegraphics[width=8.6cm]{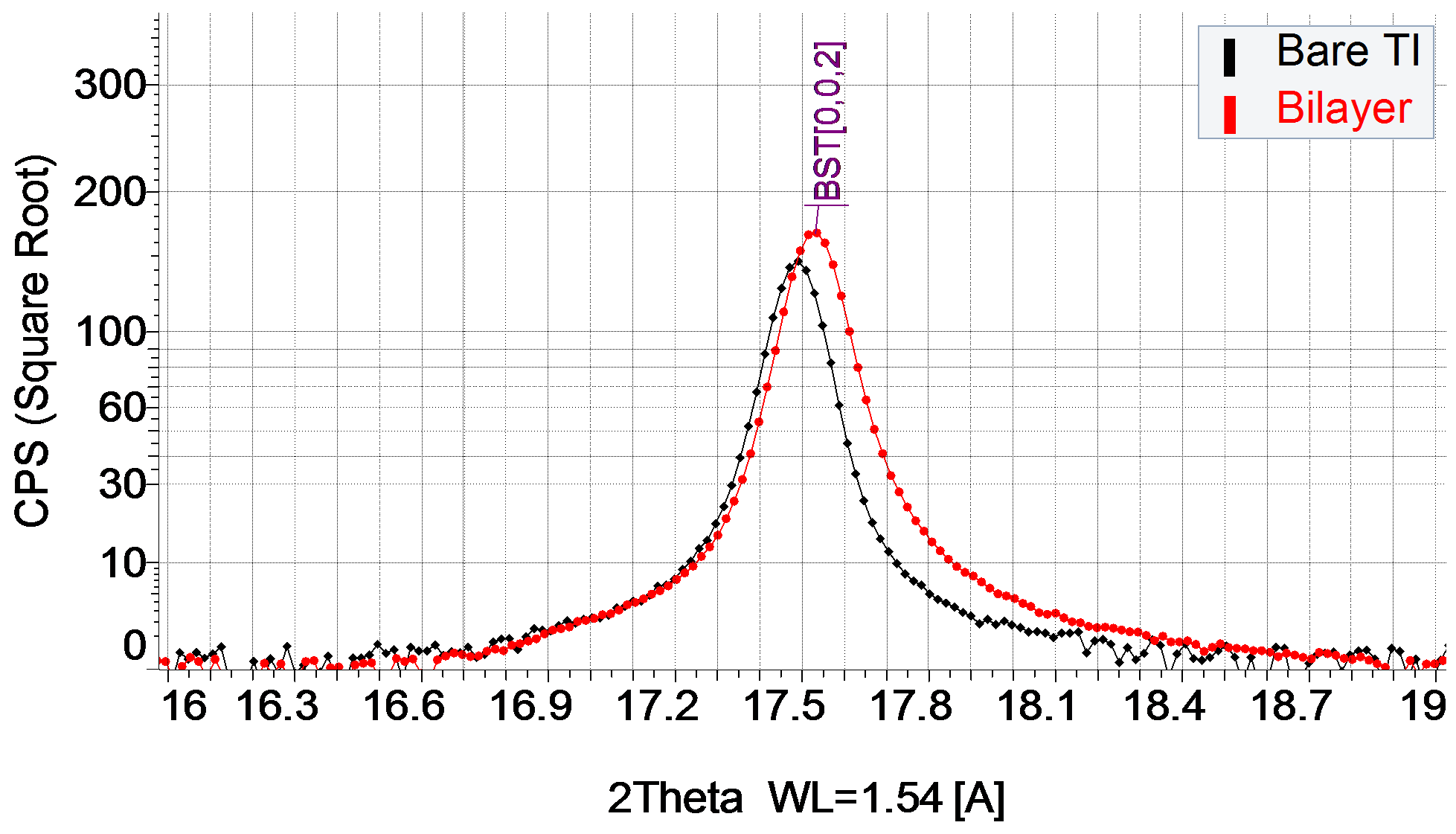}}

}

\caption{(a) X-ray diffraction data in a $\theta-2\theta$ method of a 102 [nm] thick bare TI film (black) and a bilayer with a 12 [nm] thick FM layer and a 102 [nm] thick TI layer (red). The y axis is the counts per second root. We subtracted the background from both samples. The X-ray wavelength is 1.54 [A]. (b) A zoom in on the TI [0,0,2] peak.}
\end{figure}

In figure \ref{fig:Xray} we show X-ray diffraction data of a bare 102 [nm] TI film (black), and a bilayer with 12 [nm] FM and 102 [nm] TI (red). Clearly, the films are C-axis oriented. The TI peaks intensities are similar, and the small difference is due to a different sample alignment and sample area.

Figure \ref{fig:Xrayzoom} shows a zoom on in the TI [0,0,2] peak. The bilayer peak is shifted by a $\Delta2\theta=0.04^{o}$ with respect to the bare TI peak. This means a tensile in-plain strain in the bilayer which shortens the TI C-axis lattice vector \cite{Liu2014}. It is known from the literature that a tensile in-plane strain enhances the surface states properties \cite{Liu2014}, and therefore this strain cannot be the cause for the reduced surface states coherence length we found in this article.

The bilayer peak is also broader, which can result from a smaller TI grain size. Using the Scherrer equation \cite{Borchardt-Ott2012} with a shape factor K of 0.9 \cite{Borchardt-Ott2012} we got a grain size of $~42 [nm]$ for the TI in the bilayer, and $~46 [nm]$ in the bare TI. This clearly shows that grain size is also not the source for reduced coherence length in the bilayers.

In addition, in the bare TI the surface states coherence length is about 500 [nm], which we extracted from its magneto conductivity as explained in the main article. This is much higher than the grain size, which is reasonable since the non-gapped surface states are protected from back-scattering by the grain boundaries \cite{Bernevig2013}. In the bilayers, the surfaces states are gapped and are not protected from back-scattering. This may explain why in the bilayer the coherence length is about $~85 [nm]$ (see main article), similar to the grain size. Note that our calculated grain size is a lower boundary on real grain size, which may be somewhat larger since the Scherrer equation does not take into account peak broadening by randomly oriented strain\cite{Borchardt-Ott2012}.

\section{The dominance of the exposed surface in our magneto-conductivity data\label{Detailed explanation appnd}}
In the main text we concluded that our measured low field magneto-conductivity (MC) is a property of the exposed surface. We justified this by the ability to fit the MC of bare TIs and bilayers using a one-surface modified HLN model\cite{Lu2014,Lu2011}, as was reported also by many previous publications\cite {Kim2011,He2011,Zhang2012,Wang2016,Bansal2012,Steinberg2011,Park2018,Li2019d}. We discussed the possible contributions of the bulk and the interface surface to the MC and concluded that these are smaller than the contribution of the exposed surface which dominate the MC. From this we concluded that the effective coherence length is reduced in the exposed surface by the magnetic layer and its surface states acquire an energy gap. Therefore, magnetic scattering on the interface cannot explain our effect alone, because there are no magnetic particles at the exposed surface (see appendix \ref{Fe doping appendix} why Fe dopant are not important in our system).

Here we wish to give another justification for our conclusion that our measured reduced MC in the bilayers is a property of the exposed surface. In the end, we will only stay with the assumption that in bare TIs, the contribution of the exposed surface states at least equals that of the interface surface. This is a reasonable assumption since the interface suffers more from strain, defects, impurities etc.

Let us first assume the opposite assumption- the surface states at the exposed surface of the TI in the bilayers remains intact, and the change occurs only at the interface surface. Let us also assume that the surface states at the two surfaces are not coupled and behaves as two separate transport channels. We can therefore treat the two surfaces as two transport channels connected in parallel. In this case the total conductance is the sum of the conductance of each surface, and so is the total magneto conductivity (MC): 
\begin{multline}
\Delta\sigma_{tot}\left(B\right)=
\sigma_{tot}\left(B\right)-\sigma_{tot}\left(0\right)=\\
\sigma_{1}\left(B\right)+\sigma_{2}\left(B\right)-\left(\sigma_{1}\left(0\right)+\sigma{2}\left(0\right)\right)=
\Delta\sigma_{1}\left(B\right)+\Delta\sigma_{2}\left(B\right)\label{eq:total MC parallel}
\end{multline}
The exposed surface is probably of a better quality than the interface surface, since the interface suffers more from impurities, strain, defects etc. Therefore, in a bare TI film we can assume that the contribution of the interface surface to the MC is less or equal to that of the exposed surface. 

In our bilayers the total MC is about fifth of its value in the bare TI films. Remember we are assuming that the magnetic layer reduces the MC only at the TI interface surface in the bilayers. Since the total MC is the sum of the MC of both surfaces, and the contribution of the interface surface is less or equal to that of the exposed surface, than the total MC in the bilayers cannot be less than half of the MC in bare TI. This is so unless either the MC of the interface surface changes sign or one of our assumptions is false.

If the MC of the interface surface changes sign, it means that the surface states magnetically induced energy gap is greater than the Fermi energy\cite{Lu2014,Lu2011}. For example, in our bilayer with 17 [nm]-thick TI layer this means a gap larger than ~300 [meV] (see the Fermi energy analysis in the main text). This value is more than three times larger than any previously reported or predicted magnetically induced gap at TI-FM interfaces, so we assume this scenario as highly improbable. Therefore, we must conclude that one of our assumptions is false. 

We used three assumptions: 1) In the bare TIs the contribution of the exposed surface states at least equals that of the interface surface. 2) The two surfaces are not coupled. 3) The FM layer affects only the interface surface. As we already explained, we see the first assumption as highly probable due the inferior quality of the interface surface\cite{He2011,Zhang2012}. 

If the second assumption is false, and the two surfaces behaves as a one coupled transport channel, then we cannot differ between the surface states at the interface and the exposed surfaces. Therefore, the magnetically induced gap which our MC data reflects is a property of all the surface states in the bilayers, including the exposed surface states. But this is exactly our claim- the surface states at the exposed surface of our bilayers are gapped. We also noted in the main text that for our thickest bilayers the surfaces can not couple because the coherence length is smaller than the sample thickness \cite{Steinberg2011}.

Therefore, we have to conclude that the third assumption is false, and the FM does affect the exposed surface and reduces its MC, namely reduces the effective coherence length of its surface states (see the discussion of figure 2). As we thoroughly explain in the main text, this can be done either by reducing the coherence length or the surface states, or by opening a magnetically induced gap.

Until now we did not mention the possible contribution of the bulk channel in this appendix. As we mentioned in the main text, the bulk might also contribute the weak anti localization component of the MC\cite{Steinberg2011,Park2018, Li2019d}. Time reversal symmetry does not protect the bulk states from back-scattering as the surface states in the bare TIs, and the bulk states have an energy dispersion of massive particles, not massless as the surface states. Therefore, it is highly plausible that their coherence length is also less or equal to that of the exposed surface states in the bare TIs. This makes all our previous claims about the interface surface states valid also for the bulk states. We cannot explain the attenuation of the MC in the bilayers to about fifth of the bare TI values if we assume that the exposed surface states remain intact. 

In the main text we explained that to assume that there is no gap at the exposed surface, we have to assume that the coherence length of its surface states is less than fifth of its value in the bare TIs. 
In appendix \ref{xray data appendix} we showed that the average grain size of the TI in the bilayers is almost the same as in the bare TIs (with~10\% difference), and the additional strain leads only to a shift of $~0.02^{o}$ in the [0,0,2] peak of the TI. We therefore claim that such drastic reduction of the coherence length is less probable than the opening of a magnetically induced energy gap at the exposed surface. Such a gap can occur only due to magnetic coupling to magnetic interface, and not due to magnetic scattering since there are no magnetic particles at the exposed surface.

\end{document}